\documentstyle[12pt]{article}
\def\pictures{y }
\ifx\pansw\pictures
\input prepictex
\input pictex
\input postpictex
\else
\fi
\newdimen\tdim
\def\stpltsmbl{\setplotsymbol (.)}

\def\hfloop{\beginpicture
    \setcoordinatesystem units <\tdim,\tdim>
    \setplotsymbol ({\tenrm .})
    \circulararc 360 degrees from 0 14 center at 0 0
    \circulararc 360 degrees from 0 15 center at 0 0
    \put{\circle*{6}} at -12 0
\endpicture}
\def\phrd{
\beginpicture
\setcoordinatesystem units <\tdim,\tdim> point at 2 0
\stpltsmbl
\setquadratic
\plot
0 0
2.5 -3
5 0
7.5 3
10 0
/
\endpicture}
\def\phdr{
\beginpicture
\setcoordinatesystem units <\tdim,\tdim> point at 2 0
\stpltsmbl
\setquadratic
\plot
0 0
3 -2.5
0 -5
-3 -7.5
0 -10
/
\endpicture}
\newcommand{\mysection}[1]{\setcounter{equation}{0}\section{#1}}

\begin{document}
\newcommand{\nc}{\newcommand}
\nc{\beq}{\begin{equation}}     \nc{\eeq}{\end{equation}}
\nc{\beqa}{\begin{eqnarray}}    \nc{\eeqa}{\end{eqnarray}}
\nc{\lsim}{\begin{array}{c}\,\sim\vspace{-21pt}\\< \end{array}}
\nc{\gsim}{\begin{array}{c}\sim\vspace{-21pt}\\> \end{array}}
\nc{\create}{\hat{a}^\dagger}   \nc{\destroy}{\hat{a}}
\nc{\kvec}{\vec{k}}             \nc{\kvecp}{\vec{k}^\prime}
\nc{\kvecpp}{\vec{k}^{\prime\prime} }   \nc{\kb}{\bf k}
\nc{\kbp}{{\bf k}^\prime}       \nc{\kbpp}{{\bf k}^{\prime\prime} }
\nc{\bfk}{{\bf k}}              \nc{\cohak}{a_{{\bf k}}}
\nc{\cohap}{a_{{\bf p}}}        \nc{\cohaq}{a_{{\bf q}}}
\nc{\cohbk}{b^*_{{\bf k}}}      \nc{\cohbp}{b^*_{{\bf p}}}
\nc{\cohbq}{b^*_{{\bf q}}}      \nc{\Cop}{\hat{C}}
\nc{\ir}{\vert\, i\,\rangle}       \nc{\fr}{\vert\, f\,\rangle}
\nc{\il}{\langle\, i\,\vert}       \nc{\fl}{\langle\, f\,\vert}
\nc{\A}{A_{\mu} (x)}            \nc{\ivr}{\vert\, 0\,\rangle^{\rm\, in}}
\nc{\ivl}{^{\rm\, in}\langle\, 0\,\vert}  \nc{\ovr}{\vert\, 0\,\rangle^{\rm
out}}
\nc{\ovl}{^{\rm out}\langle\, 0\,\vert}  \nc{\I}{{\rm in}}
\nc{\out}{{\rm out}}          \nc{\PA}{{\cal P} (A)}
\nc{\SA}{S^{\mu}_{\epsilon} } \nc{\JPS}{J^\mu\left(x\,\vert\,\epsilon\right)}
\nc{\ffdool}{{\rm Tr}\, F\,\widetilde{F}}
\vskip .5in
\begin{titlepage}
\begin{center}
{\hbox to\hsize{October 1994 \hfill JHU-TIPAC-940017}}
{\hbox to\hsize{hep-ph/9410407   \hfill HUTP-A0/036}}
\vskip 1 in
{\Large \bf Anomalous Violation of Conservation Laws } \\[0.125in]
{\Large \bf in Minkowski Space} \\
\vskip 0.5in

\begin{tabular}{cc}
\begin{tabular}{c}
{\bf Thomas M. Gould\footnotemark[1]}\\[.05in]
{\it Department of Physics and Astronomy}\\
{\it The Johns Hopkins University}\\
{\it Baltimore MD 21218 USA}\\[.15in]
\end{tabular}
&
\begin{tabular}{c}
{\bf Stephen D.H. Hsu}\footnotemark[2]\\[.05in]
{\it Lyman Laboratory of Physics} \\
{\it Harvard University} \\
{\it Cambridge  MA 02138 USA} \\[.15in]
\end{tabular}
\end{tabular}

\vskip .5 in
{\bf Abstract}\\[-0.05in]
\smallskip
\begin{quote}

We consider the  evolution of quantum fermi fields in
Minkowski gauge field backgrounds. Our motivation is
the study of anomalous fermion number violating processes.
We derive selection rules for fermion scattering amplitudes
which relate the violation of fermionic charge to the
change in winding between early and late times of
the vacuum part of the gauge field.
We find that the amount of fermion number
violation is always integer, even when the
topological charge of
the gauge field is fractional.
As an explicit example,
we apply our results to spherically symmetric
Minkowski solutions.

\end{quote}
\end{center}
\footnotetext[1]{gould@fermi.pha.jhu.edu}
\footnotetext[2]{hsu@hsunext.harvard.edu;
Address after January 1995:
Yale University, Sloan Physics Laboratory,

\hspace{0.2cm}Department of Physics, New Haven CT 06511.}
\end{titlepage}

\renewcommand{\thepage}{\arabic{page}}
\setcounter{page}{1}
\mysection{Introduction }

Nearly 15 years ago, Christ investigated the time evolution of
a quantum Fermi field in the background of a Minkowski space
Yang--Mills gauge field with non--vanishing topological charge~\cite{NC}.
He found that fermions are created and destroyed in accordance with
the anomalous divergence of the fermion number current,
if the gauge field reduces to a pure gauge at early and late times.
Further,
he found that when the large time limit of the gauge field contains
radiation,
the change in fermionic charge differs from what is implied by naive
integration of the anomaly equation.
The integrated anomaly equation must be supplemented by some integrals of
local polynomials of the gauge field.

In the present paper,
we reconsider the issue of anomalous fermion number violation in
Minkowski gauge field backgrounds, concentrating specifically on the
case of backgrounds which contain radiation.
Our conclusions are similar to Christ's in that we conclude that
the amount of fermion number violation is always integer,
regardless of the topological charge of the background field.
However, we generalize his analysis in several ways.
In particular,
we refine the asymptotic conditions on gauge fields necessary
to differentiate between cases with fermion number violation and
no fermion number violation.

We will refer to a ``canonical'' class of vacuum to vacuum backgrounds
with integer topological charge, which are the main subject of study
in \cite{NC}.   A more precise definition will be given in Section 5.
In agreement with \cite{NC}, we find that these configurations lead to
integer fermion number violation.
We will develop similar results for backgrounds which contain ``radiation'',
or non-pure gauge components in the far past or future.
Relaxing the canonical condition allows backgrounds with fractional
(non-integer) topological charge.
Such configurations clearly do not have a simple interpretation in terms
of the naive anomaly relation.

Our central result is that fermion number violation
is not determined
by the topological charge of the gauge background,
but rather by
the change in winding number between early and late times
of the {\it vacuum} part of the gauge background, which is
always integer.
An important observation is that
configurations in Minkowski space can have nonzero topological
charge without leading to a change in vacuum winding.

We derive selection rules which relate this quantity
to the eigenvalues of canonical fermion counter
operators acting on the asymptotic fermion in-- and out--states.
One of our selection rules applies to background configurations which
approach zero sufficiently rapidly in time in the far past and future
(precisely how rapidly will be specified in section 4).
For such backgrounds,  we have the result:
\beq
\label{srule1}
\left(\, Q_f ~-~ Q_i \,\right)~ S_{fi} ~=~ 0 ~,
\eeq
where $Q_{f,i}$ are the eigenvalues of a suitably defined counter operator
$:\!\Cop_{f,i}\! : $ applied to the final or initial states
$\fr,\ir$, and $S_{fi}$ is the $S$-matrix element
between the initial and final states:
$S_{fi}  = \fl\, i\,\rangle$.
The counter operator for in--states is
\beq
:\!\Cop_{i}\!: ~=~ \sum_{n}
\hat{a}^{\,{\rm in}\dagger}_n\,\hat{a}^{\,{\rm in}}_n ~-~
\hat{b}^{\,{\rm in}\dagger}_n\,\hat{b}^{\,{\rm in}}_n
\eeq
and $:\!\Cop_{f}\! :$ is defined similarly.
In the following section,
we give more precise definitions of all of these quantities.
The definition of $:\!\Cop\! :$ is particularly subtle since an infinite
subtraction is
required to relate the normal--ordered operator $:\!\Cop\! :$ to the naive
integrated charge operator
\beq
\label{naive}
\int d^3 x ~ J_0 (x) ~=~ \int d^3 x~ \Psi^{\dagger} (x)\, Q\,\Psi (x) ~.
\eeq
As we shall see, the anomaly appears in the dependence of the
c-number subtraction constant on the background field $A_{\mu} (x)$.

The selection rule (\ref{srule1}) implies that {\it no} anomalous
fermion number violation accompanies the class of backgrounds
which vanish sufficiently rapidly at asymptotic times (exactly how rapidly
will be discussed in section 4).
In section 5 we also derive selection rules relevant to gauge
backgrounds which lead to fermion number violation,
including both Christ's canonical case and the canonical case
with additional radiation.

Our motivation for this investigation stems from interest in the behavior
of fermion number violating amplitudes in the electroweak theory.
These processes have been studied at low energies using Euclidean techniques,
suitable for tunneling problems~\cite{BNV}.
However, at energies larger than the height of the barrier separating
n-vacua with different winding number,
we expect that fermion number violating processes will be classically allowed,
and hence describable in terms of Minkowskian paths in the functional integral.
So, our investigation is immediately relevant to some recent developments
in the study of nonperturbative processes:

\medskip
{\bf 1.} {\em Functional integral formulation for scattering amplitudes}

Gould, Hsu and Poppitz have recently developed a new formulation for
multiparticle scattering in the semiclassical approximation~\cite{GHP}.
This addresses a long standing problem in weakly coupled
gauge theories~\cite{BNV}.
The resulting formulation provides an algorithm for
constructing stationary points of the functional integral
for non-exponentially suppressed amplitudes.
In this formulation,
the stationary phase approximation to a scattering amplitude involving
many gauge particles in the final state
\beq
\label{scattamp}
\langle\, f\,\vert\, \vec{p},\, -\vec{p}\,\rangle ~,
\eeq
yields an initial value problem for real classical Minkowski trajectories.
Such amplitudes are not accompanied by the exponential WKB suppression
factor and describe processes which occur above any
energy barriers in configuration space.
The incorporation of fermions into this picture requires an understanding
of the behavior of solutions to the Dirac equation in  real Minkowski gauge
field backgrounds.

\medskip
{\bf 2.} {\em Classical solutions with fractional topological charge}

Farhi, Khoze and Singleton have recently discovered a new class of
spherically symmetric Minkowski solutions in Yang--Mills theory
{}~\cite{FKS,FKRS}.
This new development extends our knowledge of $O(3)$
symmetric solutions beyond the well--known solutions of
higher symmetry~\cite{AFF,LS}.

However, the behavior of fermions in these backgrounds has
not previously been understood because their topological charge
\beq
\label{topocharge}
\frac{g^2}{16\pi^2} \int d^4x~\ffdool ~,
\eeq
is not necessarily an integer.
The naive expectation is that this quantity determines the amount
of fermion number violation in the gauge field background,
\beq
\label{fnv}
Q_f ~-~ Q_i ~=~ \frac{g^2 N}{16\pi^2}\int d^4x~\ffdool~,
\eeq
by integrating the anomaly equation relating (\ref{topocharge})
to the divergence of the fermion number current.
This property has posed a problem for the interpretation of fermion number
in these backgrounds~\cite{FKS,FKRS}.
A major result of this paper is an understanding of the evolution of
quantum fermion fields in these non--singular non--self-dual backgrounds.

\medskip


It should not be surprising that Minkowski configurations can
have non-integer values of the topological charge.
An integer topological charge is expected only if
a configuration allows a topological classification.
This is the case for configurations which interpolate between pure
vacua, $A_i = \frac{i}{g}\,\Omega\,\partial_i \Omega^\dagger$ for
all $\vec{x}$, at early and late times $t$,
provided the gauge function $\Omega$ approaches a constant independent
of direction at large $\vert\,\vec{x}\,\vert $
\beq
\label{Ulimit}
\Omega (\vec{x}) \rightarrow 1 \hspace{1cm}
{\rm when}\hspace{1cm}\vert\,\vec{x}\,\vert\rightarrow\infty ~.
\eeq
These elements of the gauge group on compactified spacetime
are classified topologically by an integer valued Pontryagin index,
or ``winding number''
\beq
\label{pont}
\nu\left(\Omega\right) ~=~
\int {d^3x\over 24\pi^2}~ \epsilon_{ijk}~{\rm Tr}~
\Omega^\dagger\partial_i \Omega~\Omega^\dagger\partial_j \Omega~
\Omega^\dagger\partial_k \Omega
\eeq
In this case,
the topological charge (\ref{topocharge}) is also an integer,
provided spatial derivatives of the gauge potential are
negligible at large $|\vec{x}|$, so that
\beq
\label{topo}
\frac{g^2}{16\pi^2} \int d^4x~\ffdool ~ =~
\nu\left(\Omega (t=+\infty\right)) ~-~
\nu\left(\Omega (t=-\infty\right) ) ~.
\eeq
Configurations which do not interpolate between vacua,
and in particular solutions of the field equations
with finite non-zero energy,
are not classified in this manner and neither
(\ref{pont}) nor (\ref{topo}) need be integer\footnote{
However, if non-pure-gauge
configurations are dissipative, and approach
a pure gauge given by a gauge function $\Omega(x)$
as $|x| \rightarrow \infty$, such that the gauge function
$\Omega (x)$ provides a map from $S^3 \rightarrow SU(2)$,
then they can still
be classified topologically. We will see that, ultimately,
it is this classification of the
{\it vacuum} part of the configuration
which is directly related
to fermion number violation.}.
Further,
if the condition (\ref{Ulimit}) is relaxed on vacuum to vacuum
configurations,
and there is no {\it a priori} justification for it
in Minkowski spacetime,
there is no topological classification of configurations
and again neither (\ref{pont}) nor (\ref{topo}) need be integer.
In general,
configurations which do not satisfy (\ref{Ulimit}) have gauge fields
which fall off only as $1/r$ at large spatial distances.
There are many arguments for neglecting such configurations \cite{JMR},
and we will also not consider them here except in the sense
of a certain order of limits used in \cite{FKS},
which we explain in section 7.

We should note that it is not necessary to appeal to the
complicated problem of fermion number violation in quantum scattering
processes for motivation.
One can simply imagine a {\em Gedanken} experiment in which
an experimenter tunes a time-dependent external gauge field.
( If the background is a classical solution,
the experimenter need only tune the initial conditions to generate
the desired background.)
The effects of this background gauge field on the fermions in the
experiment can be considered the subject of this paper.

The remaining sections of this paper are organized as follows.
In section 2,
we define the quantities to be studied.
In section 3,
we show how the c-numbers that arise from normal ordering the anomalous
current can be first expressed as vacuum expectations of
Heisenberg operators, and then computed using ordinary perturbation theory.
This allows us to make contact with the original anomaly computations of
Adler, Bell and Jackiw~\cite{ABJ}.
We then use the results of this section to derive our first selection rule.
In section 4,
we discuss the conditions required of the background field for the validity
of the perturbative methods in section 3.
In section 5,
we study the asymptotic behavior of  {\it solutions} to the field equations
in Lorentz gauge, and show that they satisfy the conditions of section 4.
However,
we argue that Lorentz gauge can only accommodate topologically nontrivial
configurations through a discontinuous patching of gauge functions,
which potentially allows for the violation of fermion number.
In section 6,
we reproduce Christ's original results for the canonical case
with our formalism,
and show how the anomalous production of fermions requires the
use of a gauge ``twisted'' basis of solutions to the Dirac equation.
We also derive a selection rule for a general type of background which
includes radiation superposed on a canonical configuration.
When combined with the results of section 5,
this selection rule is sufficiently general to describe the fermion number
violation which results from {\it any} exact solution to the Yang--Mills
equations.
In section 7,
we review the properties of the  solutions of Farhi, Khoze, Rajagopal and
Singleton and derive an appropriate selection rule, showing that
they do not lead to fermion number production.
In section 8, we present our conclusions.

\mysection{Specifying the Problem}
Our purpose in this section is to give precise meaning to the various
mathematical quantities we will be using in this paper.
We work in the context of quantum fermi fields $\Psi (x)$ which interact
with classical background gauge fields $\A$.
We consider a system of fermions in a finite box of spatial size $L$,
which will eventually be taken to infinity in a way explained
at the end of this section.

In general, such models will exhibit anomalous divergences in global currents
$J^\mu$,
\beq
\label{anom}
\partial_\mu J^\mu ~=~  \frac{g^2 N}{16\pi^2}~\ffdool ~,
\eeq
where $N$ is the number of fermion gauge multiplets in the current.
The relation (\ref{anom}) is an operator equation, and allows the replacement
of the operator $\partial\!\cdot\! J$ with the c-number on the right hand
side of (\ref{anom}) in any matrix element.
The current is defined by
\beq
\label{JJ1}
J^{\mu} \equiv
\lim_{\epsilon \rightarrow 0}\, J^\mu\left(x \,\vert\,\epsilon\right)  ~+~
\left\{ {\bf c.t.} \right\} ~,
\eeq
where the gauge invariant, point split current is given by
\beq
\label{ps}
J^\mu\left(x \,\vert\,\epsilon\right) ~\equiv~
\overline{\Psi} (x + \epsilon/2)\,\gamma^\mu\, Q\,
{\cal P}\exp{\left(ig\int_{x - \epsilon/2}^{x + \epsilon/2} dy
\cdot A (y) \right)}
\,\Psi (x - \epsilon/2) ~.
\eeq
Here $A_\mu = T^a A^a_\mu$.
In chiral models like electroweak theory,
the generator $T^a$ includes a projection onto say
left--handed fermions.
${\cal P}$ denotes path ordering, and
$\epsilon$ is a four--vector with infinitesimal components.
We may choose $\epsilon_0 > 0$ so that (\ref{ps}) is
time-ordered.
In models such as QCD, where the fermions couple in a vectorlike way to
the gauge fields,
the anomalous current in question will be axial vector
$Q \propto \gamma_5$.
In a chiral model like the electroweak theory,
the charge matrix $Q$ is trivial in Lorentz space.
We will refer to the charge in both cases as fermion number,
although in the vectorlike case it is more properly referred
to as axial--fermion number or chirality.
Note that $\left[Q,T^a\right] = 0$, so that the position of
$Q$ in (\ref{ps}) is irrelevant.
We shall write the path ordered exponential in (\ref{ps}) more compactly
as $\PA$.

The counterterm denoted by $\{c.t.\}$ in (\ref{JJ1}) denotes the infinite
subtraction that must be made to relate the renormalized operator
$J^{\mu} (x)$ to the infinite operator
$\lim_{\epsilon \rightarrow 0} J^\mu\left(x\,\vert\,\epsilon\right)$.
As an alternative to (\ref{JJ1}), we can consider defining
\beq
\label{JJ2}
J^{\mu}~\equiv~\lim_{\epsilon\rightarrow 0}
\,\left[\, J^\mu\left(x\,\vert\,\epsilon\right) ~+~
\left\{ c.t. \right\}_{\epsilon} \,\right] ~,
\eeq
where now the counterterm is a function of $\epsilon$, becoming infinite as
$\epsilon \rightarrow 0$. The counterterm can be chosen so as to cancel the
c-number which arises from normal
ordering $J^\mu\left(x\,\vert\,\epsilon\right)$
in a {\it trivial} background $\A \equiv 0$. In section 3,
we show how to compute such c-numbers directly in perturbation theory.

The ABJ anomaly arises from the
fact that the c-number relating $J^\mu\left(x\,\vert\,\epsilon\right)$
and its normal ordered equivalent is dependent on the background field $\A$.
This dependence arises when we choose to expand the field operators
$\Psi\left(x \pm \epsilon/2\right)$
appearing in (\ref{ps}) in terms of mode functions
which are solutions to the Dirac equation in the $\A$ background.
Thus, our definition of normal ordering depends on the background.
More explicitly, we write the fermionic field operators as a
(discrete for finite $L$) sum of modes
\beq
\label{inmodes}
\Psi(x) ~=~ \sum_n\,
\hat{a}^{\rm\, in}_n\,\Psi^{\rm\, in,+}_n ~+~
\hat{b}^{\, in\dagger}_n\,\Psi^{\rm\, in,-}_n ~,
\eeq
or
\beq
\label{outmodes}
\Psi(x) ~=~ \sum_n\,
\hat{a}^{\rm out}_n\,\Psi^{{\rm out},+}_n ~+~
\hat{b}^{{\rm out} \dagger}_n\,\Psi^{{\rm out},-}_n ~.
\eeq
The operators $\hat{a}^{\,{\rm in,out}}_n$ annihilate
in-- or out-- fermions and the operators $\hat{b}^{\,{\rm in,out}\dagger}_n$
create in-- or out-- anti--fermions, respectively.
The above operators define in-- and out--vacua through the
requirements that
\beqa
\label{vacua}
\hat{a}^{\rm\, in}_n\,\vert\, 0\,\rangle^{\rm\, in} ~=~ 0
\hspace{0.5cm} & {\rm and}&  \hspace{0.5cm}
\hat{b}^{\rm\, in}_n\,\vert\, 0\,\rangle^{\rm\, in} ~=~ 0~,  \\
\hat{a}^{\rm\, out}_n\,\vert\, 0\,\rangle^{\rm\, out} ~=~ 0
\hspace{0.5cm} & {\rm and}&  \hspace{0.5cm}
\hat{b}^{\rm\, out}_n\,\vert\, 0\,\rangle^{\rm\, out} ~=~ 0  \nonumber ~.
\eeqa

The spinor functions $\Psi^{\, in,\pm}$ and $\Psi^{out,\pm}$
are orthonormal sets of functions which satisfy the Dirac equation
\beq
\label{Dirac}
\left(\,\partial\hspace{-2.5mm}/ ~-~ igA\hspace{-2.5mm}/ \,\right)
\Psi^{{\rm\, in,out},\pm}_n(x) ~=~ 0
\eeq
and boundary conditions in the asymptotic past and future respectively,
\beqa
\label{DBC}
\Psi^{\, {\rm in},\pm}_n(x)~\rightarrow~
\psi^{\pm}_n(\vec{x})~e^{\mp iE_nt}
&{\rm as}& t\rightarrow -\infty ~,  \\
\Psi^{{\rm out},\pm}_n(x)~\rightarrow~
\psi^{\pm}_n(\vec{x})~e^{\mp iE_nt}
&{\rm as}& t\rightarrow\infty \nonumber ~.
\eeqa
Here $\psi^{\pm} (\vec{x})$ are eigenfunctions of the free Dirac
Hamiltonian with positive- or negative-energy eigenvalues $\pm E_n$.
These functions $\psi^{\pm} (\vec{x})$ are complete and we choose them
to be orthonormal functions of $\vec{x}$.

Note that the boundary conditions given by (\ref{DBC}) are free fermion
boundary conditions and can not always be imposed
on the solutions of (\ref{Dirac}), specifically if the gauge field does not
vanish at asymptotic early and late times. We will be interested
in dissipative
configurations which (at least in some gauges, like
$A_0 = 0$ gauge) approach a time independent
pure gauge background
$A_i\equiv(i/g)\,R(\vec{x})\partial_i R(\vec{x})^\dagger$.
In this
case the boundary conditions in (\ref{DBC}) can be modified by including
time independent ``twists'':
\beqa
\label{TBC}
\Psi^{\,{\rm in},\pm}_n(x) &\rightarrow&
R_{\rm\, in}(\vec{x})~\psi^{\pm}_n(\vec{x})~ e^{\mp iE_nt} \nonumber \\
\Psi^{{\rm out},\pm}_n(x) &\rightarrow&
R_{\rm\, out}(\vec{x})~\psi^{\pm}_n(\vec{x})~ e^{\mp iE_nt}.
\eeqa
As will be emphasized in section 6, the necessity of these
``twists'' is central to the phenomena of fermion number violation.
Backgrounds $\A$ which allow the implementation of free boundary
conditions (\ref{DBC}) at {\it both} early and late times will be seen to
conserve
fermion number.

Substituting (\ref{inmodes},\ref{outmodes}) into (\ref{ps}) results in an
operator which is not normal ordered: it contains $\hat{b}$
operators to the left of $\hat{b}^{\dagger}$ operators. Normal ordering
the operator is equivalent to subtracting a c-number equal to the infinite sum
\beq
\label{NOsum}
S^{\mu}_{\epsilon} \left[A\right] ~\equiv~
\sum_n \, \overline{\Psi}^{-}_n (x + \epsilon/2)\,\gamma^\mu\, Q\,\PA\,
\Psi^{-}_n (x - \epsilon/2) ~,
\eeq
where the sum is over either in- or out- modes depending on whether
the operator is being normal ordered with respect to the $\ivr$ or
$\ovr$ vacuum. Thus,
\beq
\label{NOJ}
\JPS ~=~ :\!\JPS\! :_{\rm\, in,out} ~+~ \SA \left[A\right]_{\rm\, in,out}.
\eeq

As mentioned previously, the background dependence in
$\SA [A]$ comes from the implicit dependence of the solutions
$\Psi^{\pm}$ on the field $\A$. In defining the regularized current
via (\ref{JJ1}), we must choose a {\it single} value for the counterterm.
We take this to be the value of (\ref{NOsum}) when the gauge field is
trivial ($\A = 0$):
\beq
\label{JJS}
J^{\mu} (x) ~=~\lim_{\epsilon \rightarrow 0}~
\left[\,  ~:\!\JPS\! :_{\rm\, in,out} ~+~
\SA \left[A\right]_{\rm\, in,out} ~-~
S_{\epsilon}^{\mu} \left[0\right]_{\rm\, in,out}  \,\right]  ~.
\eeq
The anomalous divergence in (\ref{anom}) arises from the finite
dependence of $J^{\mu}(x)$ on $\A$ which is uncancelled by the
counterterm.

The decomposition (\ref{NOJ}) is gauge invariant in the following sense.
If under a gauge transformation
\beq
\label{AGT}
\A ~\rightarrow~ A'_{\mu} (x) = \Omega\,\A\,\Omega^\dagger ~-~
\frac{i}{g}\,\partial_i\,\Omega\,\Omega^\dagger,
\eeq
the fermion modes are also transformed
\beq
\label{PSIGT}
\Psi^{\pm}_n ~\rightarrow~ \Psi^{' \pm}_n ~=~ \Omega (x)~ \Psi^{\pm}_n,
\eeq
then each term on the right hand side of (\ref{NOJ}) remains
individually gauge invariant.
In particular, we define operators which count in- or out-modes
\beq
\label{counter}
:\! \Cop \!:_{\, in, out} ~\equiv~ \lim_{\epsilon\rightarrow 0}\int d^3x ~:\!
J^0 (x\vert \epsilon )\! :_{\rm\, in,out}~.
\eeq
These counter operators are invariant under (\ref{AGT}) and (\ref{PSIGT}),
although the modes that they count are changed.
This definition is quite natural since if the $\Psi^{\pm}_n$ are
initially chosen to be solutions of the Dirac
equation in the $\A$ background they will continue to be solutions
(now with gauge rotated boundary conditions) of the Dirac equation
in the new gauge background $A'_{\mu} (x)$.

Our general strategy for evaluating the in- and out- counters
$:\! \Cop\! :_{\rm\, in, out}$ will be to find a gauge in which the
c-numbers $\SA \left[A\right]_{\rm\, in,out}$ are easily computed.
This will typically be a gauge in which the background $A'_{\mu} (x)$ falls
off sufficiently rapidly to allow the use of standard perturbation theory, as
described in section 3.

\ifx\pansw\pictures
$$
\beginpicture
\setplotarea x from -65 to 65 , y from -40 to 55
\ellipticalarc axes ratio 3:1 360 degrees from 57 100 center at 0 100
\ellipticalarc axes ratio 3:1 -180 degrees from 80 70 center at 0 70
\ellipticalarc axes ratio 3:1  50 degrees from 80 70 center at 0 70
\ellipticalarc axes ratio 3:1 -50 degrees from -80 70 center at 0 70
\ellipticalarc axes ratio 3:1 180 degrees from -80 10 center at 0 10
\ellipticalarc axes ratio 3:1 180 degrees from -57 -20 center at 0 -20
\setdots
\ellipticalarc axes ratio 3:1 180 degrees from 80 70 center at 0 70
\ellipticalarc axes ratio 3:1 180 degrees from 80 10 center at 0 10
\ellipticalarc axes ratio 3:1 180 degrees from 57 -20 center at 0 -20
\setsolid
\putrule from 80 10 to 80 70    \putrule from -80 10 to -80 70
\arrow < 4 pt> [0.4,1.5] from 0 130 to 0 150  \put{$t$} at 0 160
\arrow < 4 pt> [0.4,1.5] from 100 40 to 120 40 \put{$r$} at 130 40
\arrow < 4 pt> [0,0] from 0 40 to 80 120
\arrow < 4 pt> [0,0] from 0 40 to 80 -40
\arrow < 4 pt> [0,0] from 0 40 to -80 120
\arrow < 4 pt> [0,0] from 0 40 to -80 -40
\putrule from -35 70 to -28 70   \putrule from 28 70 to 35 70
\putrule from -35 10 to -28 10     \putrule from 28 10 to 35 10
\putrule from 90 70 to 100 70   \put{$T_f$} at 110 70
\putrule from 90 10 to 100 10    \put{$T_i$} at 110 10
\put{{\bf I}} at 0 0 \put{{\bf II}} at 60 40 \put{{\bf III}} at 0 90
\put{$\vec{\Sigma}$} at -100 40
\put{ Figure 1: Spacetime region with spatial boundary $\vec{\Sigma}$.}
at 0 -80
\put{The ratio of the radii of cylinder and cone reflects our order of
limits.} at 20 -100
\endpicture
$$
\bigskip
\else\fi

In the original works of Adler, Bell and Jackiw~\cite{ABJ},
the anomalous divergence resulting from the sum
(\ref{NOsum}) is computed within perturbation theory.
In section 3,  we will show how the sum over exact Dirac solutions can
be expressed perturbatively,
making contact with the earlier approaches.
However, for our purposes it is crucial that the normal--ordered part of
the current be retained.
In the vacuum matrix elements studied by \cite{ABJ},
this part of the current is always null.
However, we are interested in amplitudes involving fermion excitations
in the initial and or final states, and the normal ordered part of the
integrated charge operator $\int d^3 x ~J^0(x)$ provides a counter
$:\!\Cop\! :_{\rm\, in, out}$ for these states.

We wish to specify our gauge field background in a spacetime region
like the one depicted in figure 1.
In order to discretize the fermionic energy levels,
we have taken the region to have finite spatial extent $L$,
and imposed periodic boundary conditions on the background and
the solutions $\psi^{\pm} (\vec{x})$.
We will require that {\it all} of $\vec{\Sigma}$,
the timelike boundary of the spacetime region in figure 1,
lie {\it outside} the forward light-cone of regions where $\A$ has its
support.
This will allow us to ignore all activity, whether fermionic
or bosonic, on this boundary, simply by invoking causality.
One can regard this condition as a restriction on the support of $\A$
and/or as a prescription for the order of limits when the timelike and
spacelike extents, $T_f - T_i$ and $L$, of the region are taken to
infinity.
The spatial extent $L$  should be taken to infinity first,
with $T_i$, $T_f$ fixed, as reflected in figure 1.
This order of limits will be important when we discuss explicit applications
in section 6.

\mysection{Perturbation Theory and Selection Rule}

We are now ready to derive one of our central results: a selection rule for
fermionic scattering amplitudes in gauge backgrounds which -- at least in
certain gauges such as $A_0 = 0$ gauge --
fall off sufficiently rapidly at early and late times.
The selection rule will show that configurations in this
class do not lead to fermion number violation.
(Configurations with nontrivial topology do not fall within this class.)
First, let us define in-- and out--fermion states by application of the
Heisenberg operators defined previously in (\ref{inmodes},\ref{outmodes}):
\beq
\label{state1}
\vert\, i\, \rangle^{\rm\, in} ~=~
\prod_{l}^{N_i} ~\prod_{m}^{\bar{N}_i} ~
\hat{a}^{\,{\rm in} \dagger}_{n_l} ~
\hat{b}^{\,{\rm in} \dagger}_{\bar{n}_m} ~  \ivr
\eeq
and
\beq
\label{state2}
^{\rm out} \langle\, f\,\vert ~=~  \ovl  ~
\prod_{l}^{N_f} ~\prod_{m}^{\bar{N}_f}
   ~\hat{a}^{\,{\rm out}}_{n_l}  ~ \hat{b}^{\,{\rm out}}_{\bar{n}_m} ~
\eeq
For every $l$ ($m$),
the label $n_l$ ($\bar{n}_m$) represents a set of numbers specifying
a one--fermion (one--anti-fermion) state.
The states (\ref{state1}) and (\ref{state2}) are clearly eigenstates
of the operators which count in- and out- modes,
$\hat{a}^{\,{\rm in}\dagger}\,\hat{a}^{\,{\rm in}}$ and
$\hat{b}^{\,{\rm in}\dagger}\,\hat{b}^{\,{\rm in}}$,
and
$\hat{a}^{\,{\rm out}\dagger}\,\hat{a}^{\,{\rm out}}$ and
$\hat{b}^{\,{\rm out}\dagger}\,\hat{b}^{\,{\rm out}}$, respectively.
The relation of these operators to the fermion number charge will be
explained below.
More complicated states which are not eigenstates of these operators
may be formed as superpositions of the states given above.

Our goal is to compute the matrix element given by
\beq
\label{SR}
^{\rm out} \langle\, f\,\vert  \,\left(\,\hat{C}_f ~-~\hat{C}_i\,\right)\,
\vert\, i\,\rangle^{\rm\, in},
\eeq
where $\hat{C}_{f,i} \equiv \int d^3x~ J^0 (\vec{x}, T_{f,i}  )$.
We choose to write $J^0 (\vec{x}, T_{f}  )$ in terms of an operator
which is normal ordered with respect to $\ovr$,
while $J^0 (\vec{x}, T_{i}  )$ is written in terms of an operator
which is normal ordered with respect to $\ivr$.
Using the definition of $J^0 (x)$ given by (\ref{JJS}), and the orthogonality
of mode functions $\Psi^{\,{\rm in},\pm}$ and $\Psi^{\,{\rm out},\pm}$,
we have
\beqa
\label{SR1}
\lefteqn{^{\rm out} \langle\, f\,\vert  \,
\left(\,\hat{C}_f ~-~\hat{C}_i\,\right)\,
\vert\, i\,\rangle^{\rm\, in} ~=} & & \\
& &  S_{fi} ~  \left\{\,
Q_f ~-~ Q_i  ~+~  \lim_{\epsilon \rightarrow 0}
\int d^3x ~  \left(\,
S^0_{\epsilon} \left[A(\vec{x},T_f)\right]_{\,\rm out} ~-~
S^0_{\epsilon} \left[A(\vec{x},T_i)\right]_{\,\rm in}
\,\right) \,\right\} ~. \nonumber
\eeqa
Here $S_{fi} ~=~ ^{\rm out}\langle\, f\,\vert\, i\,\rangle ^{\rm\, in}$
is the S-matrix amplitude between the chosen in- and out- states,
and $Q_{f,i}$ are the eigenvalues of the normal ordered part of the counter
operator:
\beq
\label{CQ}
:\!\Cop_f\! :\,\vert\, f\,\rangle^{\,{\rm out}} ~=~
Q_f\,\vert\, f\,\rangle^{\,{\rm out}} ~, \hspace{1cm}
:\!\Cop_i\! :\,\vert\, i\,\rangle^{\,{\rm in}} ~=~
Q_i\,\vert\, i\,\rangle^{\,{\rm in}} ~.
\eeq

We have defined $Q_{f,i}\propto\left( N_{f,i} - \bar{N}_{f,i}\right)$
for the in-- and out-- states,
with $Q_{f,i}$ taking on integer values when the charge $Q$
appearing in (\ref{ps}) is an integer~\footnote{
We assume for simplicity that this is the case.
This can be easily generalized.}.
In what follows,  we will develop perturbative methods for computing
the c-number functions $\SA \left[A\right]$ which appear in (\ref{SR1}).
We will show that when the perturbative methods apply,
\beq
\label{div}
\lim_{\epsilon \rightarrow 0}~\partial_{\mu} \SA [A] ~=~
\frac{g^2 N}{16 \pi^2}~\ffdool ~.
\eeq
Using
(\ref{div}) we can rewrite the integral in (\ref{SR1}) as
\beq
\label{in}
\lim_{\epsilon \rightarrow 0}\,\int d^4x~\partial_0 S^0_{\epsilon}
{}~=~  \frac{g^2 N}{16\pi^2}\int d^4x~\ffdool ~+~
\lim_{\epsilon \rightarrow 0}\,\int d \vec{\Sigma} \cdot
\vec{S}_{\epsilon}\left[A\right] ~,
\eeq
where as discussed in section 2, the boundary conditions on the
background $\A$ at the surface $\vec{\Sigma}$ are such that the last
integral vanishes by causality~\footnote{Our calculation is performed
in a fixed volume as described in section 2,
so four--integrals like the topological charge are well--defined at each step.
Since the topological charge eventually  cancels in our selection rule
(\ref{SR3}),
we need not be concerned with whether the integrals in (\ref{in}) continue
to be well--defined when we take the volume to infinity.
For a discussion of this possibility, see \cite{5authors}.}.

We can also apply the operator relation (\ref{anom}) directly to the matrix
element (\ref{SR}). This yields
\beq
\label{SR2}
 ^{\rm out} \langle\, f\,\vert  \int d^4x~ \partial_0 J^0 (x)
\,\vert\, i\,\rangle^{\rm\, in}
{}~=~  \left(\, \frac{g^2 N}{16\pi^2}
\int d^4x~\ffdool \,\right) ~S_{fi} ~+~
^{\rm out} \langle\, f\,\vert\,
\int d \vec{\Sigma} \cdot \vec{J}\,\vert\, i\,\rangle^{\rm\, in} ~,
\eeq
where again the last integral over $\vec{\Sigma}$ is zero.
Equating (\ref{SR2}) and (\ref{SR1}) then yields the desired selection rule:
\beq
\label{SR3}
\left(\, Q_f ~-~ Q_i\,\right)\, S_{fi} ~=~ 0 ~.
\eeq
Thus, there is no fermion number violation in backgrounds for which
the perturbative methods apply.

A simple way to understand this result is to note that in the break--up of
$J^{\mu} (x)$ into normal ordered plus c-number terms the anomaly
was fully accounted for in the c-number part.
Therefore, the divergence of
the normal ordered part of $J^{\mu} (x)$ is precisely zero:
\beq
\label{DNJ}
\partial_{\mu} :\! J^{\mu}(x)\! : ~=~ 0 ~.
\eeq
This is sufficient to guarantee fermion number conservation in the class of
backgrounds for which this relation holds.
In section 4,
we will see that for backgrounds with nontrivial topology the above
arguments are not valid,
primarily because we cannot use ``untwisted'' bases of in- and out-
solutions to the Dirac equation which are
necessary for the derivation of equation (\ref{div}). When ``twisted''
fields are used, the analog of (\ref{div}) has no anomalous term on the
RHS, and the effect of the anomaly is manifested in a nonzero difference
between the eigenvalues of the initial and final counter operators,
requiring the RHS of (\ref{DNJ}) to be nonzero.

In the remainder of this section,
we will show how the c-number functions
$\SA [A]$
can be computed in perturbation theory,
culminating in the expression (\ref{div}).
We first express $\SA [A]$ directly in terms of
vacuum matrix elements:
\beq
\label{vev}
\SA\left[A\right]_{\rm\, in,\, out} ~=~ \ovl ~
\overline{\Psi} (x + \epsilon/2)\,\gamma^\mu\, Q \,{\cal P}(A)
\,\Psi (x - \epsilon/2)~\ivr ~.
\eeq
It is clear that the normal ordered part
of the current $~:\!\JPS\! :~$
is annihilated when it acts
either to the right
(if we normal order with respect to in--states)
or to the left (if we normal order with respect
to out--states).
Thus what remains in the above matrix element is just
$\SA[A]_{\rm\, in}$, which therefore is identical to $\SA[A]_{\rm out}$.

To compute the matrix element in (\ref{vev})
using perturbation theory,
we first go to the Interaction picture.
Until now our discussion has been in
the Heisenberg picture, with
field operators evolving in time as
$\Psi(\vec{x},t) ~=~ U (t,0) ~\Psi(\vec{x},0)~U^\dagger(t,0)$,
where the time evolution operator for the time dependent Hamiltonian
is $U(t',t)\equiv \exp{i\int^{t'}_t dt'' H(t'')}$.
The vacuum and all other states are time independent.
We denote fields in the Interaction picture by $\phi(t)$, with
\beqa
\label{Ipic}
\phi(\vec{x},t)^{\rm\, in, \, out} &\equiv &
e^{iH_0 t}~U^\dagger (t,0)~\Psi(\vec{x},t)^{\rm\, in,\, out}~
U(t,0)~e^{-iH_0 t} \\
&=& U_I(t,\pm\infty) ~ \Psi(\vec{x},t)^{\rm\, in, \, out} ~
U_I^{\dagger}(t,\pm\infty)
\nonumber ~.
\eeqa
Here $H_0$ is the free Hamiltonian, and the evolution operator
$U_I(t',t) ~=~ \exp{-i\int^{t'}_t dt'' H_I(t'')}$ is given in terms of the
interacting part of the Hamiltonian.
The (time dependent) Interaction picture vacuum is given by
\beq
\label{Ivac}
\vert\, 0,t\,\rangle ~=~ U_I (t, -\infty)\,\ivr ~=~ U_I (t,+\infty)\,\ovr ~.
\eeq

Substituting (\ref{Ipic}) into the matrix element (\ref{vev}) yields
\beqa
\label{vev1}
\lefteqn{\SA [A] ~=} & & \\
& & \langle\, 0, t\,\vert\, U_I^{\dagger} (-\infty,+\infty)~
{\cal T}\left(\,
\bar{\phi} (x+\epsilon/2)~\gamma^{\mu}~Q~{\cal P}(A)~\phi
(x-\epsilon/2)~U_I(t,-t)
\,\right)\vert\, 0, -t\,\rangle \nonumber ~.
\eeqa
Expanding the exponential $U_I(t,-t)$
inside the time ordered product
reproduces ordinary perturbation
theory in terms of free Feynman
propagators and insertions of the background $\A $.
The factor of
$U_I^{\dagger} (-\infty,+\infty) =
\exp +i \int_{-\infty}^{+\infty} dt'\, H'_I (t')$
divides out disconnected graphs in the usual way.

One can now explicitly evaluate (\ref{vev1}) using standard Feynman rules.
Diagrammatically, (\ref{vev1}) corresponds to the sum of graphs depicted in
figure 2.
To derive (\ref{div}), which is the divergence $\partial_{\mu} (\cdots)$
of the matrix element (\ref{vev}), we simply note that the corresponding
graphs are precisely those computed in the usual perturbative derivation
of the anomaly~\cite{ABJ}.
\ifx\pansw\pictures
$$
\beginpicture
\setcoordinatesystem units <\tdim,\tdim>
\setplotarea x from -40 to 60 , y from -45 to 70
\stpltsmbl
\put {$S^\mu_\epsilon (A) ~=$} [r] at -30 0
\multiput {\hfloop} at 0 0   115 0  60 -80 /
\multiput {\phrd} at 142 0 *2  10 0 /
\put {$+$} at 70 0 \put {$+$} [l] at 200 0
\startrotation by 0.66667 -0.74536 about 0 0
\multiput {\phdr} at 110 26 *2 0 10 / \stoprotation
\startrotation by -0.66667 -0.74536 about 0 0
\multiput {\phdr} at 10 132 *2 0 10 / \stoprotation
\put {$+$} at 30 -80  \put {$+$} [l] at 150 -80 \put {$\cdots$} at 180 -80
\put {Figure 2: Perturbative expansion of $S^\mu_\epsilon$ includes
fermion loops} [l] at -80 -140
\put {with one insertion of $J^\mu$ and any number of external gauge
bosons.} [l] at -80 -160
\endpicture
$$
\else\fi

\mysection{Sufficient conditions on $A_{\mu}$}

In this section,
we discuss the conditions on $\A$ which allow the use of
perturbation theory in the calculation of (\ref{vev}).
The issue here is the use of the Interaction picture, which
assumes that the Heisenberg field
$\Psi (x)$ approaches the free field $\phi (x)$ at asymptotically early and
late times.
This requires that the $U$-matrix defined in (\ref{Ipic}) satisfy
the condition
\beq
\label{Ucon}
\lim_{t \rightarrow \pm \infty} U_I(t,\pm \infty) ~=~ 1 ~,
\eeq
as well as a similar condition with the
arguments of $U$ reversed. This condition
requires that $\A$
fall off sufficiently rapidly to allow
the use of
free asymptotic states as
in- and out- states at early and late times.
The physical implication is that the same (free fermion) basis of
solutions can be used in the canonical quantization of the field at
both early and late times. This implies
that the in- and out- vacua
are the same (both are annihilated by creation operators of the
interaction-picture field $\phi^I (x)$) and hence there is no fermion
number violation, as seen explicitly in the selection rule derived
in the previous section for backgrounds of this type.

To study the behavior of the $U$-matrix in an arbitrary background, we will
use the Yang--Feldman equation which relates the Heisenberg field to a
perturbative expansion in terms of the free field boundary condition,
the free propagator, and the background $\A$. We write
\beq
\label{YF}
\Psi_n (\vec{x},x_0 ) ~=~ \phi_n (\vec{x},x_0) ~+~ \int d^4 y~
\Delta (x-y)\, A_{\mu}(y)\,\gamma^{\mu}\,\Psi_n (y) ~,
\eeq
It is straightforward to check that $\Psi_n (x)$ defined in this way
satisfies the Dirac equation in the
$\A$ background.
The choice of free propagator $\Delta (x-y)$
determines the boundary conditions that $\Psi_n (x)$ satisfies.
For an advanced propagator  $\Delta (x-y)$
with support only for $y_0 > x_0$, (\ref{YF}) implements
the advanced boundary condition that
$\Psi_n (\vec{x},x_0) \rightarrow \phi_n (\vec{x},x_0)$ as
$x_0 \rightarrow + \infty$,
provided that the integral on the RHS of the equation approaches zero
in this limit.
If the integral does not vanish in that limit,
the Dirac equation in the $\A$ background is incompatible with free
boundary conditions in the far future.
Similar considerations apply for
retarded boundary conditions and a
retarded propagator $\Delta (x-y)$.
We will study the case of advanced
boundary conditions below.

Our goal here is to determine what conditions on $\A$
are sufficient for the integral in (\ref{YF}) to vanish.
We first expand the integral perturbatively
by using (\ref{YF}) recursively. This yields
\beq
\label{YFE}
\Psi_n (\vec{x},x_0) ~=~ \phi_n (\vec{x},x_0) ~+~
\int d^4 y~\Delta (x-y)\, A_{\mu}(y)\,\gamma^{\mu}\,\phi_n (y) ~+~\cdots
\eeq
where the ellipsis indicate an infinite sum of multiple integrals.
In the case of massless fermions, the propagator takes the form
\beq
\label{AP}
\Delta (x-y) ~=~ \partial_{\mu}\,\left[\,
\theta (y_0 - x_0) ~ \delta ((x-y)^2)\,\right]\, \gamma^{\mu}
\eeq
with support only on the forward lightcone $(x-y)^2 = 0$, $y_0 > x_0$.

Given the asymptotic behavior of the background field,
we can bound the Yang--Feldman integrals in (\ref{YFE}).
For large times, we will take
\beq
\label{AB}
\vert\,\A\,\vert ~,~ \vert\,\partial_\nu\,\A\,\vert ~<~ B (x_0)
\hspace{1cm}\forall~\vec{x} ~,
\eeq
where the positive definite function $B(x_0) \rightarrow 0$ as $t \rightarrow
\infty$.
(Since $\vert\,\A\,\vert $ and  $\vert\,\partial_\nu\A\,\vert $
have different dimensions,
we really mean two different functions $B(x_0)$ and $B'(x_0)$ in (\ref{AB}).
But for notational simplicity we will refer only to one function,
related to the other by some factor with units of mass.)
Additional conditions on $B(x_0)$ (i.e. that it approach zero
appropriately rapidly) will be specified below and will depend on specific
properties of the background $\A$.

For free fermion solutions, $\vert\phi_n (x)\vert $ is constant at large times
\footnote{It is usual in scattering theory to think of the in- and out-
states as wavepackets, which by the Riemann--Lebesgue lemma
vanish exponentially far from where they are localized. Were this the
case in this analysis the exponential fall-off in
$I_{YF}$ would lead to the very weak requirement that
$B(x_0) \rightarrow 0$. However, wave packet states are too restrictive
for our purposes here. That is, we wish
our counter $\Cop$ to be defined over a complete set of
fermion modes, even those
which are very close to pure plane waves or have spatial extent
of order the size of our ``box''.
To this end, we must study
the Yang--Feldman integral with boundary conditions given by plane wave
functions $\phi_n (x)$. This yields much stronger conditions on $\A$.}
and
we can write
\beqa
\label{YFI2}
I_{YF} &=&   \int d^4 y~ \Delta (x-y) A_{\mu} (y)\gamma^{\mu}\phi_n (y) \\
&<&  C \int_{x_0}^{\infty} dy_0~(y_0 - x_0)~B(y_0) \nonumber  ~.
\eeqa
where $C$ is a numerical constant.
$I_{YF}$ vanishes at large times if
\beq
\label{AB1}
B(x_0) ~<~ \frac{D}{x_0^2}
\eeq
at large times for some constant $D$.
A more sophisticated analysis of $I_{YF}$ in (\ref{YFI2}) leads to less
stringent requirements on $B(x_0)$.
In particular, if one considers the oscillatory behavior of $\phi_n (y)$,
it appears that the integral vanishes for $B(x_0) < D/x_0$,
although this requires more detailed analysis.
However it is clear that without some additional
information about the asymptotic behavior of $\A$
the weak condition that $B(x_0) \rightarrow 0$ is insufficient to guarantee
vanishing of $I_{YF}$.
In \cite{NC}, the condition
\beq
\label{NCcondition}
\lim_{t \rightarrow \infty} ~ A_{\mu} (\vec{x},t) \rightarrow 0
\eeq
was given as sufficient for a type of perturbative analysis. It is easy
to see that there are backgrounds which satisfy (\ref{NCcondition})
but for which the Yang--Feldman integrals do not vanish at asymptotic times.
The condition (\ref{NCcondition}) does not appear sufficient to
justify subsequent computations in \cite{NC}.

We can derive a more restrictive bound than (\ref{AB1}),
in the case of the spherically symmetric solutions of~\cite{FKS}
and in general for the Lorentz gauge classical solutions
to be discussed in section 5.
As we will see in more detail in sections 5 and 7,
the solutions $A_{\mu} (y)$ only have support on or near the lightcone
($y^2 = 0$) at late times
and furthermore retain their ``shape'' in a soliton--like
fashion along the lightcone.
We can then see from (\ref{AP}) that the Yang--Feldman
integral $I_{YF} (x)$ has its support restricted to the forward lightcone
of the point $x_{\mu}$: $(x - y)^2 = 0$.
The intersection of the two lightcones given by $(x-y)^2 = 0$ and $y^2 = 0$
is

\smallskip
{\bf a.} $|\vec{x}| < x_0$~:  the null set.

{\bf b.} $x^2 = 0$~: a null line given by $y_{\mu} \propto x_{\mu}$.

{\bf c.} $|\vec{x}| > x_0$~: a one-dimensional surface given the solution
$y$ to the equation ($\hat{y} = \vec{y}/ |\vec{y}|$):
\beq
\label{LC}
y_0 ~=~ {x^2\over 2\,\left(x_0 - \vec{x} \cdot \hat{y}\right)} ~.
\eeq
Note that in case {\bf c.}, as $y_0 \rightarrow \infty$,
the angle between $\hat{x}$ and $\hat{y}$
approaches a constant such that $\hat{x} \cdot \hat{y} ~=~ x_0 / | \vec{x} |$.

For cases {\bf b.} and {\bf c.},
the Yang--Feldman integral is of the form
\beq
\label{SAYF}
I_{YF} ~=~ C \int_{x_0}^{\infty} dy_0~ f\left(y_0\right)~
\exp\left(\, i\,\sqrt{2}\, p_n^{\mu}\cdot\hat{l}_{\mu}\, y_0\,\right) ~,
\eeq
where $p_n^{\mu}$ is the four--momentum of the
plane-wave function $\phi_n\left(y\right)$ and $f\left(y_0\right)$
is a function whose norm is asymptotic to $B\left(y_0\right)$.
$\hat{l}_{\mu}$ is a unit vector from $x_{\mu}$ along its forward lightcone,
which
in case {\bf c.} is itself a function of $y_0$ that approaches a constant
only at large $y_0$.
The right hand side goes to zero as $x_0 \rightarrow \infty$ if
$B\left(y_0\right) \leq D/ y_0^k$ for some constant $D$ and $k > 0$.
One can check that this is also sufficient for the vanishing of the subsequent
integrals in (\ref{YFE}).
As we will see in sections 5 and 6, this condition is satisfied by all Lorentz
gauge solutions and also by the solutions of Farhi et. al. in the
so-called $\varphi = 0$ gauge.

For backgrounds satisfying (\ref{AB}), with
appropriately strong conditions on $B(x_0)$, the vacuum expectation values
(\ref{vev}) can be reduced to standard perturbation theory
computations in the interaction
picture. Our selection rule therefore applies to any such
background $\A$.
In fact, the selection rule follows directly from
the existence of a $U$-matrix satisfying (\ref{Ucon}).
The existence of such a $U$-matrix implies that both
Heisenberg fields $\Psi^{\,{\rm in}} (x)$ and $\Psi^{\,{\rm out}} (x)$
approach the {\it same}
free field $\phi^I (x)$ in the asymptotic past and future
respectively.
This implies that the in- and out- vacua are {\it identical} from
the standpoint of the fermions,
and hence that there can be no charge violation.

It is worth noting that a canonical gauge background $\A$ which has
nontrivial topology will in general fall off no faster than
$1/ \vert\, x_0\,\vert $ at large times, regardless of gauge,
and have support in regions of finite solid angle,
as opposed to only on the lightcone.
We can understand this in the following way.
Suppose that a
configuration $\A$ reduces to a pure gauge function
on the compactified surface of spacetime $S_3$.
Then, $\A$ is given on that surface in terms of a gauge function
$\Omega (\theta_i)$ ($\theta_i$ are angular variables.).
If the configuration is topologically nontrivial,
$\Omega (\theta_i)$ must provide
a nontrivial map $\Omega : S_3 \rightarrow SU(2)$, and hence
$\partial_{\theta} \Omega \neq 0$ in some region of solid angle.
It is easy to see that the resulting gauge fields in this
solid-angular region fall off no faster than $1/\vert\, x_{\mu}\,\vert $.
Such configurations do not allow the construction of a time evolution
$U$-matrix satisfying (\ref{Ucon}),
and therefore require ``twisted'' boundary conditions in {\it any} gauge.
In other words, they require us to define {\it different} fermionic
in- and out- vacua at early and late times.
In this sense,
topologically nontrivial configurations can be said to interpolate
between different vacua.
Configurations which possess (possibly fractional) topological charge,
but which fall off sufficiently rapidly to allow identical fermionic
in-- and out--vacua, can be said to remain in
the same vacuum throughout their time evolution.

\mysection{Asymptotics of Lorentz gauge solutions}

In this section, we show that Minkowski Yang--Mills solutions to the
Lorentz gauge classical equations of motions are sufficiently
well-behaved for the application of the results of section 3.
In particular,
we will show that at asymptotic times these solutions have their
support only on or near the lightcone and furthermore fall off at least
as fast as $1/\vert\, t\,\vert$.
The first part of the result should come as no surprise,
since excitations in pure Yang--Mills theory travel only at the speed
of light.

In Lorentz gauge, $~\partial_{\mu} A^{\mu} (x) = 0~$, the
linearized equations of motion
reduce to the free wave equation for each degree of freedom.
Therefore, at very early or late times, we can write any solution in
the form
\beq
\label{LoSol}
A_{\mu} ( x ) ~=~ \int d^3 k~  A_\mu (\vec{k}) ~ e^{i k \cdot x} ~+~ \{ {\rm
h.c.} \},
\eeq
where the
four--vector $k_{\mu} = \left(\pm\vert\,\vec{k}\,\vert, \vec{k}\right)$
and $k_{\mu} A^{\mu} (\vec{k}) = 0$.
We will assume that the momentum space function $A_\mu (\vec{k})$
is smooth and integrable, with the same being true of its first few
derivatives.

Now consider the large $x_0$ behavior of the above integral.
Let us consider the behavior of the gauge potential along
rays given by $\vert\,\vec{x}\,\vert ~=~ a\, x_0$,
where $a$ is some constant.
We will show that for $a \neq 1$ the resulting $A_\mu ( x_0 )$
falls off at least as fast as $1/\vert x_0\vert^2$,
and that for $a = \pm 1$ the
falloff is at least as fast as $1/\vert x_0\vert$.
This is sufficient to satisfy the
conditions derived from the Yang--Feldman equations in the
previous section.

We first write (\ref{LoSol}) in spherical coordinates
\beq
\label{LoS1}
A_\mu\left(x_0\right) ~=~
\int dk~ d \phi~ \int_{-1}^{1} du~k^2~
A_\mu\left(k,\phi, u\right) ~
e^{i k\, x_0\,\left(\, au \,\pm\, 1\,\right)} ~+~ \{ {\rm h.c.} \}  ~.
\eeq
Next, integrating by parts with respect to $u$ gives
\beqa
\label{LoS2}
\lefteqn{
\int_{-1}^{1} du~A_\mu\left(k, \phi, u\right)~
e^{i k\, x_0\,\left(\, au\,\pm\, 1\,\right)} ~+~
\left\{ {\rm h.c.} \right\} ~=} & &  \\
\hspace{1cm}&=& \frac{1}{ika x_0} \int_{-1}^{1} du~
A_\mu\left(k, \phi, u\right)~
\partial_u~e^{i k\, x_0\,\left(\,au\,\pm\, 1\,\right)} ~+~
\{ {\rm h.c.} \}  \nonumber \\
\hspace{1cm}&=&
{A\left(k,\phi,u \right)\over ika x_0}~
e^{i k\,x_0\,\left(au\,\pm\, 1\right)} \Bigl|^{\,1}_{-1} ~-~
\frac{1}{ika x_0 }\,\int_{-1}^{1} du~\partial_u A_\mu\left(k,\phi, u\right)
{}~e^{i k\, x_0\,\left(\, au\,\pm\, 1\,\right)} ~+~
\{ {\rm h.c.} \}  \nonumber ~.
\eeqa
By the Riemann--Lebesgue lemma
and the assumption of smoothness of $\partial_u A$,
the last integral approaches zero
as $x_0 \rightarrow \infty$, so the leading asymptotic behavior of the
$du$ integral is given by the first term in the last equation of (\ref{LoS2}).
In fact, by repeating the same process of partial integration
on the last integral, we can show
that corrections to the leading behavior go to zero at least as fast as
$1/ |x_0|^2$.

Reinserting the result of (\ref{LoS2}) into (\ref{LoS1}), we get
\beqa
\label{LoS3}
\lefteqn{
A_\mu (x_0) ~=} & & \\
& &  \int dk~ d \phi~ \left[\,  e^{ik x_0 (a \pm 1)} \,
\frac{A_\mu(k,\phi,u=1)}{ikax_0}
 ~-~ e^{ik x_0 (- a \pm 1)}\, \frac{A_\mu(k,\phi,u= - 1)}{ika x_0}
 ~+~ \{ {\rm h.c.} \} \,\right] ~+~ \cdots \nonumber ~,
\eeqa
where the ellipsis denote terms which, after integration, fall off faster than
$1/ |x_0|^2$.
For $a = \pm 1$ (i.e. along a lightcone),the integral contains terms
which fall off as $1/| x_0 |$ or faster. However,
for $a \neq 1$ we can again repeat the partial integration procedure (this time
in the variable $k$) to conclude that
the entire integral falls off faster than $1/ |x_0|^2$.

We recall from the previous section that sufficient conditions for the
convergence of the Yang--Feldman expansion were that the background
fall off like $1/ |x_0|^2$ in regions of finite solid angle, or like
$1/| x_0 |$ if its support is localized on the lightcone.
We have shown here that essentially any solution to the Lorentz gauge
equations of motion posesses this property and hence does not lead to the
violation of fermion number.

How general is Lorentz gauge? To use the Lorentz gauge
condition both at early and late times, it is necessary to know whether
it is always possible to gauge transform a given configuration to
Lorentz gauge globally.
(One can always go to Lorentz gauge in a local region.)
Unfortunately, we suspect that at least some
topologically nontrivial
configurations cannot be tranformed to the Lorentz gauge via a
continuous gauge function $\Omega (\vec{x},t)$.
A similar result has been demonstrated
by Jackiw, Munizich and Rebbi \cite{JMR} for the
Coulomb gauge ($\partial_i A^i = 0$). In the Coulomb gauge, the classical
vacuum, which satisfies $F_{\mu \nu} = 0$, is uniquely $\A = 0$.
This in turn implies that topologically nontrivial configurations,
such as trajectories connecting nontrivial vacua in the $A_0 ~=~ 0$ gauge
{\it without} the Coulomb condition, can only be accommodated with the
Coulomb condition by means of a discontinuous gauge transform.
Essentially, it is necessary to discontinuously
patch together two distinct gauge functions in order to accommodate
the nontrivial topology in the more restrictive Coulomb gauge.

If Lorentz gauge is similar to Coulomb gauge in that it
cannot accommodate
configurations which interpolate between
distinct vacua without a discontinuity in
gauge, then
we cannot using the asymptotic results
from the earlier part of this section to prove the
nonexistence of fermion number violating Minkowskian
solutions.
However, we expect that up to an additional gauge
transform at either early or late times (which may
be discontinuous if extended into the entire spacetime),
any classical solution
can be made to have asymptotic behavior which allows the use
of the perturbative methods of section 3. This observation
will allow us to apply the more general of the selection
rules we derive in the next section to any classical solution.

\mysection{Canonical case revisited}

In this section,
we wish to study backgrounds $\A$ in which fermion number
is actually violated.
We will first consider the canonical configurations studied by
Christ~\cite{NC}, and then proceed to a hybrid case which also contains
some additional ``radiation'' and which may therefore have non-integer
topological charge.
Again we will see that the fermion number violation is integral.

Fermion number violation can be seen as a consequence of a
failure of the condition
(\ref{Ucon}) or equivalently of the free fermion boundary conditions
in (\ref{DBC}).
If $\A$ does not fall off asymptotically, the conditions (\ref{DBC}) are
inconsistent with the Dirac equation.
However, if $\A$ approaches a pure gauge at large times,
we can instead use the ``twisted'' boundary conditions given by (\ref{TBC}).
Realistic finite energy configurations must be of this type since there are
no non-disipative solutions to the pure Yang--Mills equations.
Since under these conditions the out-- and in--modes (and therefore the out--
and in--vacua) are in general different, the relevant out-- and in--counters
can register different total charges in the future and past.

Let us first consider the canonical configurations studied
by Christ~\cite{NC}.
These configurations interpolate between time independent, pure gauge
configurations in the far past and far future,
and have integer topological charge.
However, in order to posess  nonzero topological charge,
$\A$ must approach a non-zero time independent configuration either
in the past or future (or both).
For simplicity in such cases,
we will assume that the gauge field is zero in the far past
and nontrivial in the far future~\footnote{We always
assume that $\Omega (\vec{x})\rightarrow 1$ as
$|\vec{x}|\rightarrow\infty$, so that its winding number is integer.}
\beq
\label{canonical}
\A ~=~ 0 \hspace{0.5cm}{\rm for}\hspace{0.5cm} x_0 ~<~ -T ~, \hspace{1cm}
A_i (\vec{x}) ~=~ - \frac{i}{g}\,\Omega (\vec{x})\,\partial_i\,\Omega
(\vec{x})^\dagger \hspace{0.5cm}{\rm for}\hspace{0.5cm}  x_0 ~>~ T ~.
\eeq
Here $T$ is some arbitrary fiducial time.
These backgrounds do not satisfy the conditions (\ref{AB}),
and do not lead to solutions of the Dirac equation consistent
with (\ref{DBC}).
Instead, the twisted conditions (\ref{TBC}) are appropriate, with
$R_{\rm\, in}(\vec{x}) = 1$ and $R_{\rm\, out}(\vec{x}) = \Omega (\vec{x})$.

Let us revisit our selection rule, but this time defining the out-states,
out-vacuum and out-counter in terms of the twisted modes defined above.
In other words, for operators in the far future, we expand
\beq
\label{TBCnew}
\Psi(x) ~=~ \sum_n \hat{a}^{\rm out}_n\,\Psi^{{\rm out},+}_{n} ~+~
\hat{b}^{{\rm out} \dagger}\,\Psi^{{\rm out},-}_{n}
\eeq
but with
\beq
\label{TBC1}
\Psi^{{\rm out},\pm}_{n} (x) \longrightarrow
U (\vec{x})\,\psi^{\pm}(\vec{x})\, e^{\mp iEt}
\eeq
in the asymptotic future.
The matrix element of interest is then
\beq
\label{TSR}
^{\rm out} \langle\, f\,\vert\, \left(
\hat{C}_f ~-~ \hat{C}_i \right) \,
\vert\, i\,\rangle^{\,\rm in}.
\eeq
We can rewrite (\ref{TSR}) as
\beqa
\label{TSR1}
\lefteqn{
^{\rm out} \langle\, f\,\vert\, \left( \hat{C}_f ~-~
\hat{C}_i \right)  \,\vert\, i\,\rangle^{\rm\, in} ~=} & & \\
& & S_{fi} ~  \left\{
Q_f  ~-~  Q_i  ~+~  \lim_{\epsilon \rightarrow 0} \int d^3x ~
\left(\,
S^{\, 0}_{\epsilon} \left[A(\vec{x},T_f)\right] ~-~
S^0_{\epsilon} \left[A = 0\right]
\,\right)~\right\} ~, \nonumber
\eeqa
where again
\beq
\label{TCQ}
:\!\Cop_{f}\! :\,\vert\, f\,\rangle^{\rm out} ~=~
Q_{f}\,\vert\, f\,\rangle^{\rm out} ~.
\eeq
Now, however, the quantity
\beq
\label{TNOsum}
S^{\, 0}_{\epsilon} \left[A\right] ~\equiv~
\sum_n  \overline{\Psi}^{\,{\rm out},-}_{n} (x + \epsilon/2)\,
\gamma^0\, Q\,\PA\,\Psi^{{\rm out},-}_{n} (x - \epsilon/2) ~,
\eeq
is a sum over the twisted, out- modes given in (\ref{TBCnew}).
This sum cannot be reduced
to an Interaction picture, perturbative calculation as before
(when the sum was over untwisted modes),
because the necessary $U$-matrix no longer exists. However, we
can calculate (\ref{TNOsum}) directly by
removing the gauge twists $\Omega (\vec{x})$ from the definition
of $\Psi^{{\rm out},-}_{n,t}$ and using them
to gauge transform the Wilson line ${\cal P} (A)$:
\beq
\label{GT}
\Omega (\vec{x})^\dagger ~ {\cal P} (A) ~ \Omega (\vec{x})  ~=~  1
\eeq
since $\A ~=~ A_i (\vec{x}) ~=~
-\frac{i}{g} \Omega (\vec{x}) \partial_i \Omega (\vec{x})^\dagger$
at large times $T_f > T$.
This leaves a sum over {\it free} modes, with no path ordered
expononential. In other words,
\beq
\label{SGT}
S^{\, 0}_{\epsilon} \left[\, A(\vec{x},T_f)\,\right] ~=~
S^0_{\epsilon} \left[\, A = 0\,\right] ~,
\eeq
where $S^{\, 0}_{\epsilon}$ on the left is a sum over
twisted modes and on the right is a sum over untwisted modes.
This implies
that the integrand in (\ref{TSR1}) is exactly zero.
Thus, using the anomaly equation as in (\ref{SR2}), we have
\beq
\label{TSR2}
\left(\, Q_f ~-~ Q_i\,\right)~S_{fi} ~=~
\frac{g^2 N}{16 \pi^2}~ \ffdool ~.
\eeq
This selection rule implies that the final state of any nonzero
scattering amplitude
must possess {\it twisted} fermion which is different
from the fermion number of the initial state, and furthermore this net
difference
must be integral.
Our selection rule is equivalent to the operator relation first
derived by Christ in this context ((3.12) in \cite{NC}).

It is worth noting that in the above calculation the anomaly did
{\it not} appear in the c-number terms which result from normal
ordering. This is of course due to our use of the gauge twisted
basis. In the new basis a net change in fermion number is necessary
to satisfy the anomaly equation.

Having reproduced the known results for the canonical type of
background,
we now derive a selection rule valid for the hybrid case of
a canonical background which also contains some additional ``radiation''
and which may therefore have non-integer topological charge.
We will again see that the fermion number violation is integral.

We consider backgrounds which in the far future are equivalent up to
a gauge transformation $\Omega (\vec{x})$
to configurations which satisfy the Interaction picture conditions of
section 4.
{}From the results of the previous section, we expect that any classical
solution falls into this category.
As usual,
we consider gauge transforms with time independent gauge function
$\Omega (\vec{x}) \rightarrow \Omega_{\infty}$
(where $\Omega_{\infty}$ is constant) as $|\vec{x}| \rightarrow \infty$.
In other words, we require that in the far future our background $\A$ satisfy
\beq
\label{MGB}
A_i (x) ~=~ \Omega\, A'_i (x)\,\Omega^\dagger ~-~
\frac{i}{g}\,\partial_i\,\Omega\,\Omega^\dagger
\eeq
where $A' (x)$ satisfies (\ref{AB}) with appropriate $B(x_0)$.
For simplicity,
we assume that in the far past the gauge field also satisfies
(\ref{AB}) for appropriate $B(x_0)$, allowing the use of perturbative
methods in the far past \footnote{Note that because of the nontrivial topology,
$\Omega (\vec{x})$ cannot be smoothly
extended to an $\Omega (\vec{x},t)$ valued
in the entire spacetime which relates $A_i (\vec{x},t)$ everywhere to an
$A'_i (\vec{x},t)$ which satisfies the conditions of Section 4.
However, this {\it can} be done for the spherically symmetric solutions of
section 7}.

We now define the out--states, out--vacuum and out--counter in terms of modes
\beq
\label{ABC}
\Psi^{\,{\rm out},\pm}_n\left(x\right) ~=~
\Omega (\vec{x})~\Psi'^{\,{\rm out}, \pm}_n \left(x\right)
\eeq
where $\Psi^{\,{\rm out},\pm}_n$ is a solution to the Dirac equation in
the $A_i (x)$ background and $\Psi'^{\,{\rm out},\pm}_n$
denotes solutions to the Dirac equation in the $A'_i\left(x\right)$ background,
with the latter satisfying out- boundary conditions as in (\ref{DBC}).
The in--modes are solutions to the Dirac equation in the $A_i\left(x\right)$
background, but satisfying the free fermion in-- boundary
conditions of (\ref{DBC}).
The field operator in the asymptotic future is expanded as
\beq
\label{ABCexp}
\Psi(x) ~=~ \sum_n
\hat{a}^{\,{\rm out}}_n\,\Psi^{\,{\rm out},+}_n ~+~
\hat{b}^{\,{\rm out} \dagger}\,\Psi^{\,{\rm out},-}_n ~.
\eeq
The matrix element of interest is then
\beq
\label{ASR}
^{\rm out}\langle\, f\,\vert\,\left(\,\hat{C}_f~-~\hat{C}_i\,\right)
\,\vert\, i\,\rangle^{\,\rm in} ~,
\eeq
where for example the state $~^{\rm out} \langle\, f\,\vert~$
is constructed by acting with creation operators in the
$\Psi^{\,{\rm out},\pm}_n$ basis acting on the $A$-vacuum --
i.e. those annihilated by all $\hat{a}^{\,{\rm out}}_n$,
$\hat{b}^{\,{\rm out}}_n$ operators in the same $A$ basis.
For notational simplicity,
we will denote the gauge field $A_{i,f} = A\left(\vec{x},T_{i,f}\right)$.

Following our now familiar procedure, (\ref{ASR}) can be rewritten as
\beqa
\label{ASR1}
\lefteqn{^{\,\rm out} \langle\, f\,\vert\,\left( \hat{C}_f ~-~
\hat{C}_i~ \right) \,\vert\, i\,\rangle^{\,\rm in} ~=} & & \\
& & \hspace{1cm} S_{fi} ~  \left\{\,
Q_f  ~-~  Q_i  ~+~  \lim_{\epsilon \rightarrow 0}
 \int d^3x ~  \left(\,
S^{\, 0}_{\epsilon} \left[ A_f\right] ~-~
S^{\, 0}_{\epsilon} \left[ A_i\right] \,\right)
\,\right\} ~. \nonumber
\eeqa
with
\beq
\label{ACQ}
:\!\Cop_{i,f}\! :\,\vert\, i,f\,\rangle^{\rm\, in,out}
{}~=~ Q_{i,f}\,\vert\, i,f\,\rangle^{\rm\, in,out} ~.
\eeq
and
\beq
\label{ANOsum}
S^{\, 0 }_{\epsilon} (A) ~=~ \sum_n
\overline{\Psi}^{\,{\rm out},-}_n (x + \epsilon/2)\,\gamma^0\, Q\,\PA\,
\Psi^{{\rm out},-}_n (x - \epsilon/2) ~.
\eeq
Again, we cannot compute $S^{\, 0}_{\epsilon}\left[A_f\right]$
in perturbation theory because $A_f$ does not approach zero
in the far future.
On the other hand, $S^{\, 0}_{\epsilon}\left[A_i\right]$
is reducible to a sum of Feynman graphs because we have chosen $\A$ to
satisfy (\ref{AB}) in the asymptotic past.

To evaluate $S^{\, 0}_{\epsilon}\left[A_f\right]$,
we must first use (\ref{ABC}) to remove the gauge transforms from the
fermion modes in (\ref{ANOsum}).  This yields
\beq
\label{ANOsum1}
S^{\, 0}_{\epsilon} (A) ~=~ \sum_n
\overline{\Psi}\,'^{\,{\rm out},-}_n (x + \epsilon/2)\,\gamma^0\,
Q\,{\cal P} (A')\,\Psi'^{\,{\rm out},-}_n (x - \epsilon/2) ~,
\eeq
where we have again used the gauge transformation property of the Wilson line
and also equation (\ref{MGB}):
\beq
\label{WGT}
\Omega (\vec{x})^\dagger ~ {\cal P} (A)~\Omega (\vec{x})  ~=~
{\cal P} (A') ~.
\eeq
The final expression (\ref{ANOsum1}) can now be evaluated in perturbation
theory
because $A'(x)$ is well-behaved in the asymptotic future.
Thus both $S^{\, 0}_{\epsilon}\left[A_i\right]$ and
$S^{\, 0}_{\epsilon}\left[A_f\right]$ reduce to the sum of
Feynman graphs depicted in figure 2,
evaluated in the $A_i$ and $A'_f$ backgrounds respectively.

The integral in (\ref{ASR1}) is then simply equal to
\beq
\label{diff}
 q\left(A'_f\right) ~-~ q \left(A_i\right) ~,
\eeq
where the Chern--Simons number is defined as
\beq
\label{q}
q\left(A\right) ~\equiv~ \int d^3x~ K^0 ~=~
{g^2\over 16\pi^2}\int d^3x~\epsilon^{0ijk}~
{\rm Tr}~\left(A_iF_{jk} ~+~\frac{2}{3}\,i\,g\,A_iA_jA_k\right) ~.
\eeq
To simplify (\ref{diff}) further, we use the following result
\beq
\label{qtransf}
q\left(A^\Omega\right) ~=~
q\left(A\right) ~+~ \nu\left(\Omega\right) ~+~
{i\,g\over 8\pi^2}\int_S d\sigma_i~ \epsilon_{ijk}~{\rm Tr}~
\Omega^\dagger\partial_j\Omega~A_k \,
\eeq
where the winding number of the map $\Omega$ is
\beq
\label{wind}
\nu\left(\Omega\right) ~=~
\int {d^3x\over 24\pi^2}~ \epsilon_{ijk}~{\rm Tr}~
\Omega^\dagger\partial_i \Omega~\Omega^\dagger\partial_j \Omega~
\Omega^\dagger\partial_k \Omega ~.
\eeq
When the map $\Omega$ is constant on the sphere $S_2$ at infinity,
the surface term in (\ref{qtransf}) vanishes.
Further, (\ref{wind}) is then integer-valued,
according to homotopy arguments. We denote the winding
of our map $\Omega(\vec{x})$ by $\nu\left(\Omega\right) = n$.

Thus, (\ref{diff}) can be rewritten as
\beq
\label{diff1}
 q (A_f) ~-~ q (A_i) ~+~ n.
\eeq
Finally, returning to (\ref{ASR1}), we use the anomaly equation as in
(\ref{SR2}) to obtain
\beq
\label{ASR3}
\left(\, Q_f ~-~ Q_i\,\right)~S_{fi} ~=~
\left[\, -n ~+~ \frac{g^2 N}{16 \pi^2}\,\ffdool
{}~-~ \left(\, q (A_f) ~-~ q (A_i) \,\right) \,\right] ~S_{fi} ~.
\eeq
The $\A$ field has no support on the spatial boundary $\vec{\Sigma}$
so we have
\beq
\frac{g^2}{16 \pi^2}\,\int d^4x~\ffdool ~=~ q (A_f) ~-~ q (A_i) ~.
\eeq

Our selection rule is finally
\beq
\label{ASR4}
\left(\, Q_f ~-~ Q_i\,\right)~S_{fi} ~=~ -n\, N\, S_{fi} ~.
\eeq

Again, the physical interpretation here is that the radiation
component of the background field -- i.e. that part of the background
which is consistent with the $U$-matrix construction as discussed in
section 3, and which therefore does not invalidate the usage of free
solutions as asymptotic states and does not contribute to fermion
number violation.
Gauge backgrounds with nonzero energy can be considered ``near''
a particular vacuum configuration if their radiation (non-gauge)
components fall off rapidly enough to satisy the
conditions derived in section 4. From the results of section 5
we know that this is true of all exact solutions.
{}From the analysis here we
see that the fermion number violation is actually controlled
by the winding number of the nearby vacuum.

\mysection{Application to Spherically Symmetric Solutions}
In this section,
we specify some of the previous results to the case of
spherically symmetric solutions to the $SU(2)$ Yang--Mills
equations~\footnote{The notation is from \cite{FKS}.}.
Recently, these solutions have received
attention~\cite{FKS,FKRS}, in part  because they may
in general yield values of the winding number and topological charge
which are not integers.
As such, their interpretation in terms of fermion number violation has
not previously been understood.

We will demonstrate that solutions of the type considered
in \cite{FKRS}, which can be developed and studied within
a perturbative expansion,
exhibit the
type of asymptotic
behavior derived more generally in section 5, and hence
do not lead to fermion number violation.
In the selection rule governing fermion number violation,
additional terms conspire to cancel the naive topological charge.
Nevertheless, these solutions continue to be interesting in the
context of multiparticle scattering, where they provide
stationary points for scattering amplitudes~\cite{GHP}.

The $SO(3)$--symmetric ansatz is given in terms of four scalar functions
of $(r,t)$: $a_0$, $a_1$, $\rho$, and $\varphi$
\beqa
\label{spherical}
A_0\left(\vec{x},t\right) &=&
{1\over 2g} ~ a_0\, {\bf\sigma}\cdot \hat{\bf x} \\
A_i\left(\vec{x},t\right) &=&
{1\over 2g}\,\left(\, a_1 \, e^3_i ~+~
{\rho\,\sin\varphi\over r}\, e^1_i ~+~
{1-\rho\,\cos\varphi\over r}\, e^2_i\,\right) ~,
\eeqa
where the matrix-valued functions $\{e^k_i\}$ are defined as
\beq
\label{matrices}
e^1_i ~=~ \sigma_i - {\bf\sigma} \cdot \hat{\bf x}\,\hat x_i~, \hspace{1cm}
e^2_i ~=~   i \left[{\bf\sigma} \cdot \hat{ \bf x}\,\sigma_i - \hat
x_i\right] ~=~ \epsilon_{ijk} \hat x_j \sigma_k~, \hspace{1cm}
e^3_i ~=~ {\bf\sigma} \cdot\hat{\bf x}\,\hat x_i ~,
\eeq
and where $\hat{\bf x}$ is a unit three-vector.
Here
$F_{\mu\nu}= \frac{1}{2}\,\sigma^a F^a_{\mu\nu} = \partial_\mu
A_\nu - \partial_\nu A_\mu - i g \left[ A_\mu, A_\nu\right]$ is the field
strength and $A_\mu= \frac{1}{2}\,\sigma^aA^a_\mu$.

The  ansatz preserves a residual $U(1)$ subgroup of the $SU(2)$
gauge group consisting of the transformations,
\beq
\label{SPGT}
\Omega(\vec{x},t) ~=~
\exp \left[\, \frac{i}{2}\, f(r,t)\, {\bf \sigma}\cdot\hat{\bf x}
\,\right] ~.
\eeq
under which the fields transform as
\beq
\label{SGT2}
a_\mu ~\to~ a_\mu ~+~ \partial_\mu f~,\hspace{1cm}
\varphi~\to~ \varphi ~+~ f  ~.
\eeq
where $\mu=0,1=t,r$. The field $\rho$ is gauge invariant.
We can construct another gauge invariant field in
terms of the $(1+1)$--dimensional field strength
\beq
\label{psi}
\psi~\equiv~\frac{1}{2}\, r^2\,\epsilon^{\mu\nu}~\partial_\mu~a_\nu ~.
\eeq
The two gauge invariant variables $\rho$ and $\psi$,
together with the gauge variant $\varphi$,
specify the spherically symmetric solution completely.

The equations of motion yield a relation for
the gauge fields $a_{0,1}$
\beq
\label{relation}
\partial^\alpha \psi ~=~
- \epsilon^{\alpha\nu}\,\rho^2\,
\left(\,\partial_\nu\varphi ~-~ a_\nu\,\right) ~,
\eeq
where $\epsilon_{01}=+1$.
Now given $\psi$ and $\rho$, one can determine $a_\mu$ and $\varphi$,
after fixing a gauge.
At sufficiently early or late times, it can be shown ~\cite{FKRS} that $\psi$
approaches
a soliton-like form with constant shape and magnitude which decreases as
$1/t$ and has support only near the lightcone. It can also be shown that
$\rho \rightarrow 1$ at sufficiently early or late times for any finite energy
solution.
This leaves a ``bounce'' region of size $\equiv \Delta$ where the solution
behaves in a nonlinear fashion.

We will need two different gauges for the discussion to follow:
$A_0=0$ gauge and $\varphi=0$ gauge.
The two are related by a spherical gauge transformation (\ref{SPGT}),
and we will make use of this relationship when we calculate the fermion
number violation in this background.
We describe these gauges below and some specific properties of the
spherical ansatz in these gauges.

\medskip
{\bf 1.}\hspace{2mm}{\em $A_0=0$ gauge}

This  condition allows a residual time independent gauge freedom.
It is equivalent to $a_0=0$, from (\ref{spherical}).
So, integrating (\ref{relation}),
we can determine the scalar field $\varphi$ up to an arbitrary
time independent function $\varphi_0\left(r\right)$.
\beq
\label{phia0}
\varphi\left(r,t\right) ~=~
-\int_{-\infty}^t dt'~
{\partial_r \psi\left(r,t'\right)\over \rho\left(r,t'\right)^2} ~+~
\varphi_0\left(r\right) ~.
\eeq
For large $r=R$, the integral gets contributions only from the
light cone regions where $\psi$ has support.
So, when $t<-R$, the integral is approximately zero.
When $-R<t<R$, the integral gets a contribution from the
past lightcone.
When $t>R$, the integral gets two contributions, one from the
past and one from the future lightcones.
Note that a vanishing $\rho\left(r,t'\right)$ can lead to a line of
singularities for $\varphi\left(r, t > t'\right)$. In the case of
the perturbative solutions of \cite{FKRS} $\rho\left(r,t'\right)$ never
vanishes, however solutions discussed earlier in
\cite{FKS} do exhibit zeros of $\rho\left(r,t'\right)$. We will restrict
our attention to the perturbative solutions for the moment.

To fix the gauge completely we set $\varphi_0=0$.
Then, we can see that $\varphi\left(r,t\right)$
can have support only near the lightcones
and in the ``bounce'' region near the origin $r, t \sim \Delta$.
Outside of the ``bounce'' region
the integrand of (\ref{phia0}) has its support only
near the lightcone region(s) at $\left(r, t' < t\right)$.
When we integrate over the lightcones at large $r$,
we can use the fact
that $\partial_r \psi\left(r,t'\right)\simeq\pm\partial_t\psi\left(r,t'\right)$
and that $\rho\left(r,t'\right) \simeq 1$ to write
\beq
\label{phia01}
\varphi\left(r,t\right)~=~\pm~
\left(\, \psi\left(r, -\infty\right) ~-~
\psi \left(r,t\right)\,\right) ~=~ \mp ~ \psi\left(r,t\right) ~.
\eeq
Thus $\varphi\left(r,t\right)$ is only nonzero near the lightcone, since
$\psi\left(r,t\right)$ only has support there at large $r$ or $t$.

However, for $r \sim \Delta$ it is easy to see that in $A_0 = 0$ gauge the
spatial components $A_i\left(x\right)$
do not necessarily approach zero at large times,
since  $\varphi\left(r \sim \Delta,t \rightarrow \infty\right)$
can approach a nonzero constant.
This behavior generically prevents our use of perturbative methods
to evaluate matrix elements in the far future.

\medskip
{\bf 2.}\hspace{2mm}{\em $\varphi=0$ gauge}

The $\varphi=0$ gauge is ``physical'', in the sense that the gauge field
is completely specified by the gauge invariant variables $\psi$
and $\rho$.
The field $a_\nu$ is determined  directly by (\ref{psi})
\beq
\label{phi0}
a_{0,1}~=~-{1\over \rho^2}~\partial_{1,0}\,\psi ~.
\eeq

The $\varphi = 0$ and $A_0 = 0$ gauges are related by
a spherical gauge transformation with
$\Omega\left(r,t\right) =-\varphi\left(r,t\right)$.
Note that
$\Omega\left(r\rightarrow\infty, t\right)\rightarrow 0$,
since in the original $A_0 = 0$ gauge
$\varphi\left(r\rightarrow\infty, t\right)\rightarrow 0$.
A ``snapshot'' at some fixed, late time of a spherically
symmetric solution in
$\varphi = 0$ gauge would reveal the following features:

\smallskip
{\bf a.} A spherical region of size $R \sim t$ in which the field is
pure gauge, and possesses nontrivial winding or angular dependence.

{\bf b.} A thin surface region, in which there is nonzero energy
density and non-pure gauge activity.

{\bf c.} The region outside the sphere where the field is pure
gauge with constant orientation.
\smallskip

In the $\varphi = 0$ gauge,
the solutions have support only near the lightcone,
and fall off sufficiently rapidly to allow the construction of a $U$ matrix
and the use of perturbative methods.
Thus, by our previous arguments,
there should be no anomalous fermion number violation in the background
of these solutions -- the selection rule of section 3 should apply.
We can also examine this case in more detail using the same analysis we
developed
for the last selection rule in section 6 (equations (\ref{MGB})
to (\ref{ASR4}) ).
This will exhibit some of the intricacies of counting fermions in different
gauges.
Here, let the gauge field $\A$ be a spherically symmetric
solutions in the
$A_0 = 0$ gauge,
while the transformed field $A'_{\mu}\left(x\right)$ is the solution in the
$\varphi = 0$ gauge.
The two configurations are related by a spherical gauge transform with
$\Omega\left(r,t\right) =-\varphi\left(r,t\right)$ which is now time-dependent.
The relevant fermion modes, which are solutions
to the Dirac equation in the $A$ and $A'$  backgrounds, are
\beq
\label{SPFM}
\Psi^{\pm}_n\left(x\right) ~=~
\Omega\left(\vec{x},t\right)~\Psi'^{\pm}_n\left(x\right) ~,
\eeq
where $\Omega\left(\vec{x},t\right)$ is defined by (\ref{SPGT}).
Note that due to the
good behavior of the $A'$ fields at
asymptotic times, we can impose
plane wave boundary conditions on the
in-- and out--modes of $\Psi^{\pm}_{A',n}\left(x\right)$, as in (\ref{DBC}).
The in-- and out--modes of $\Psi^{\pm}_{A,n}(x)$ obey similar
boundary conditions,
but now twisted by the gauge function $\Omega\left(\vec{x},t\right)$.

The corresponding selection rule in the spherical ansatz case then
follows from repeating the steps from (\ref{ASR}) to (\ref{ASR4}), with the
gauge function $\Omega\left(\vec{x},t\right)$
replacing $\Omega\left(\vec{x}\right)$.
Since $\Omega\left(\vec{x}\rightarrow\infty,t\right)\rightarrow 1$,
the surface integral in (\ref{qtransf}) is again zero. This leaves
\beq
\label{SASR}
\left(\, Q_f ~-~ Q_i \,\right)~S_{fi} ~=~ \nu\left( \Omega\right)~S_{fi}~.
\eeq
The winding number of the map $\Omega $ can be evaluated
(using (5.4) from \cite{FKS}), but for our order of limits where
$r \rightarrow \infty$ first, with $T_f$ and $T_i$ held fixed,
\beq
\label{FKSwind}
\nu\left(\Omega\right) ~=~ \frac{1}{2\pi}\,
\left[\, f\left( r = \infty, T_f\right) ~-~
\sin  f\left( r = \infty, T_f\right)\,\right] ~.
\eeq
Here we have used the fact that $~f( r = 0, t) ~=~ 0~$.
Since, as noted above,
$f\left( r = \infty, T_f\right) = 0$ also,
we see that there is no fermion number
violation, in agreement with our previous more general argument.

It is important to recognize that the order of limits we use here
is different from that in Farhi, Khoze and Singleton~\cite{FKS}.
These authors found generally non--integer values for the winding number
and charge and speculated about their interpretation in terms
of ``fractional violation of fermion number''.
In our formulation however,
the amount of fermion number violation associated with their solutions
is always integral,
and in particular for those of \cite{FKRS} is always zero.
The spacetime region we consider has finite temporal extent, from
$T_i$ to $T_f$, with spatial extent $L$ which should be taken to infinity
first, with $T_i$ and $T_f$ fixed.
Therefore, the quantities which appear in our analysis are integrals of
the gauge field over an infinite three-volume, but at a fixed time.
In order to compute what Farhi et al. define as winding number,
they first take the limit $t \rightarrow \infty$, thereby reducing their
configuration to a pure gauge, and then subsequently perform a spatial
integral.
In the $\varphi = 0$ gauge ``snapshot'' described above, this corresponds
to integrating only over the region {\bf a}, or inside the sphere where
the field is a pure gauge.

The analysis of this section leading to the conclusion that fermion number
is not violated, strictly only applies to solutions
which have $\rho\left(r,t\right) \neq 0$ at all times.
This is the case for the solutions found in \cite{FKRS},
which are constructed as perturbations around the vacuum $\rho=1$
at all times.
Solutions which approach distinct vacua at early and late
times, which we expect to  result in integer fermion number violation,
necessarily require at least one point where $\rho \rightarrow 0$.
One can see this by recalling that the spherical ansatz reduces
$SU(2)$ Yang--Mills theory to a $\left(1+1\right)$ Abelian Higgs model
in curved spacetime,
with a complex Higgs scalar $\chi\equiv -i\rho\,\exp{i\varphi}$.
Configurations in the Abelian Higgs model can only have nontrivial
topology if the norm of the complex scalar
field $\vert\,\chi\,\vert= \rho$ vanishes.
Our previous analysis does not necessarily apply to such configurations,
due to the possible singularities induced in both the $A_0 = 0$ and
$\varphi = 0$ gauge descriptions of such configurations.

An explicit example of this type is a particular case of the
well--known L\"uscher--Schechter solutions~\cite{LS}.
These solutions have the gauge invariant function
\beq
\label{LS}
\rho\left(r,t\right)^2 ~=~ 1 ~+~ q\left(q ~+~ 2\right)\,
\cos^2 w\left(r,t\right) ~,
\eeq
where $q\left(\,\tau=\tau\left(r,t\right)\,\right)$
describes a ``particle'' in a double well potential,
\beq
\label{doublewell}
\ddot q ~+~ V^\prime(q) ~=~ 0 ~, \hspace{1cm}
V\left(q\right)~=~{1\over 2}\, q^2\left( q ~+~ 2\right)^2 ~,
\eeq
and $\dot q \equiv dq/d\tau$.
Since $0\le\cos^2w\le 1$, and $q\left(q+2\right)\ge -1$,
$\rho$ can vanish only if $q=-1$ and $\cos^2w=1$.
This corresponds to a shell in spacetime $(r,t)$.

Now, solutions to (\ref{doublewell}) with non-constant $q$ fall into
two classes,
depending on whether $\varepsilon \equiv {1\over 2} \dot q^2 + V(q)$
is smaller or larger than $1/2$, the barrier height of $V$ at the unstable
point $q=-1$.
For $\varepsilon< 1/2$,
the ``particle'' oscillates in either well as a function of $\tau$,
never reaching $-1$, and $\rho$ never vanishes.
So, our previous analysis applies and fermion number is not violated.
However, for $\varepsilon<1/2$,
the ``particle'' oscillates up and over the barrier at $q=-1$,
and $\rho$ vanishes at one point in spacetime $(r,t)$.
This solution approaches a distinct vacua at early and late
times, and would result in integer fermion number violation.

\mysection{Conclusions}

We have shown that the naive expectation that fermion number
violation is determined solely
by the topological charge of the background gauge field is
incorrect. Whether fermion number
is violated in a given background depends on other
factors -- primarily on whether the fermionic
in- and out- vacua determined by the asymptotic
behavior of the background field are different.
One can make the distinction between
different vacua in terms of what set of asymptotic
boundary conditions
are compatible with the solutions
of the Dirac equation. This in turn leads to conditions
on the rate at which the background field
must fall off asymptotically.

Backgrounds which as $|x_{\mu}| \rightarrow \infty$
approach a pure-gauge potential  determined
by a gauge function satisfying
$~\Omega (|\vec{x}| \rightarrow \infty) ~\rightarrow~ \Omega_{\infty}~$
can be classified topologically.
This includes any dissipative configurations,
and in particular all exact solutions.
Those that have nontrivial topology in this sense
fall off too slowly
(in {\it any} smooth gauge) to allow free
fermion boundary conditions at both
early and late times.
We have shown that such backgrounds lead
to integer fermion number violation.
It is important to note that configurations in Minkowski
space can have nonzero topological charge without being
topologically nontrivial in the sense defined above.

Our most general selection rule (\ref{ASR4}) applies to
any background which is a solution to the Yang--Mills equations
of motion (subject to some smoothness conditions on the initial or
final data).
The amount of fermion number violation is given by
$$
\left(\, Q_f ~-~ Q_i\,\right)~S_{fi} ~=~ -n\, N\, S_{fi} ~,  \nonumber
$$
in terms of the change in winding $n$ of the pure--gauge or vacuum part
of the configuration.  $n$ is always integer--valued.
In section 5,
the non-gauge part of any classical solution was shown
to fall off
rapidly enough to allow our analysis and to not contribute to
any fermion number violation.

The spherically symmetric solutions discussed in Section 7 provide an example
of configurations with nonzero topological charge but which do {\it not}
lead to any fermion number violation.
For these configurations,
it is always possible to find a gauge in which the potential falls off
rapidly enough to allow free fermion boundary conditions at both early
and late times.
While both topologically nontrivial configurations
and the spherically symmetric solutions of section 7
can by suitable choice of gauge be made to fall
off like $~ 1 / |x_0| ~$ at asymptotic times,
the crucial difference is that the nontrivial
configurations will have support in a region of nonzero
solid angle,
whereas the spherical symmetric solutions (in $\varphi = 0$ gauge)
have their support localized on the lightcone.
This results in very different
behavior of the relevant Yang--Feldman integrals
as discussion in Section 4.

The usual heuristic picture of the energy landscape of
configuration space is drawn with energy
plotted versus Chern--Simons number.
It is clear from our investigation that
Chern--Simons number alone is not sufficient to characterize,
even in a coarse way, the location of a configuration
in this space. It is possible
to construct configurations with nonzero,
noninteger Chern--Simons number which are
nevertheless identical to the trivial
vacuum from the viewpoint of the
fermi levels and fermionic vacuum. From
this point of view, the solutions of
\cite{FKRS} do not pass ``over the barrier'' in any sense,
but rather remain in the
basin of the trivial vacuum.

We have restricted ourselves here to the analysis
of classical gauge fields. However in
a formulation which includes quantized gauge fields --
i.e. a functional integral over both gauge and fermion trajectories --
our selection rules will apply to each gauge trajectory in the measure.
The selection rules therefore determine which semiclassical paths lead to
fermion number violation.
The results of \cite{GHP} show that the existence
of classical, Minkoskian solutions that lead to fermion number violation
implies unsuppressed fermion number violating amplitudes in the
theory with quantized gauge fields (i.e. resulting from the scattering
of gauge bosons). The question that remains to be answered is
whether there exist exact solutions with nontrivial topology
which reduce at early times to wave packets of the type
which can be produced at accelerators -- e.g. consisting
of a small number of high energy particles.

\newpage
\centerline{\bf Acknowledgements}
\vskip 0.1in
The authors would like to thank Edward Farhi, Valya Khoze,
Krishna Rajagopal and Bob Singleton for useful discussions.
TMG acknowledges the support of the National Science
Foundation under grant \mbox{NSF-PHY-90-96198} and
\mbox{NSF-PHY-94-04057.}
SDH acknowledges support from the National Science Foundation
under grant \mbox{NSF-PHY-87-14654,}
the state of Texas under grant \mbox{TNRLC-RGFY-106,}
the Harvard Society of Fellows and an SSC Fellowship.
SDH would also like to thank the Institute for Theoretical Physics at UCSB
and the Lawrence Berkeley Laboratory for their hospitality while some of
this work was performed.

\nc{\ib}[3]{        {\em ibid. }{\bf #1} (19#2) #3}
\nc{\np}[3]{        {\em Nucl.\ Phys. }{\bf #1} (19#2) #3}
\nc{\pl}[3]{        {\em Phys.\ Lett. }{\bf #1} (19#2) #3}
\nc{\pr}[3]{        {\em Phys.\ Rev.  }{\bf #1} (19#2) #3}
\nc{\prep}[3]{      {\em Phys.\ Rep.  }{\bf #1} (19#2) #3}
\nc{\prl}[3]{       {\em Phys.\ Rev.\ Lett. }{\bf #1} (19#2) #3}

\end{document}